\documentclass[twocolumn,10pt]{article}
\usepackage[utf8]{inputenc}
\usepackage[margin=1in]{geometry}
\usepackage{graphicx}
\usepackage{caption}
\usepackage{titlesec}
\usepackage{titling}
\usepackage{charter}
\usepackage{wrapfig}
\usepackage[normalem]{ulem}
\usepackage{lineno}
\usepackage[normalem]{ulem}

\titleformat{\section}[hang]{\large\bfseries}{\thesection}{0.5em}{}

\titleformat{\subsection}[hang]{\normalfont\itshape}{\thesubsection}{0.5em}{}

\usepackage[colorlinks,bookmarksopen,bookmarksnumbered,citecolor=blue,urlcolor=blue,linkcolor=blue]{hyperref}
\usepackage{color}
\usepackage{authblk}
\usepackage{bm}
\usepackage{amsmath,amssymb}
\usepackage{moreverb}
\usepackage{comment}
\usepackage{natbib}
\newcommand\blfootnote[1]{%
  \begingroup
  \renewcommand\thefootnote{}\footnote{#1}%
  \addtocounter{footnote}{-1}%
  \endgroup
}
\providecommand{\keywords}[1]
{
  \small	
  \textbf{\textit{Keywords---}} #1
}
\title{Buoyancy-Driven Entrainment in Dry Thermals}

\author[1]{Brett McKim\thanks{a}}
\author[2]{Nadir Jeevanjee}
\author[3]{Daniel Lecoanet}
\affil[1]{University of California, Santa Barbara, Department of Physics}
\affil[2]{Princeton University, Department of Geosciences}
\affil[3]{Princeton University, Princeton Center for Theoretical Science}

\date{August 2019\thanks{a}}

\begin{document}
% \linenumbers
\twocolumn[
  \begin{@twocolumnfalse}
    \maketitle
    \begin{abstract}
\citet{turner1957} proposed that dry thermals entrain because of buoyancy (via a constraint which requires an increase in the radius $a$). This however, runs counter to the scaling arguments commonly used to derive the entrainment rate, which rely on either the  self-similarity of \citet{scorer1957} or the turbulent entrainment hypothesis of \citet{morton1956}. The assumption of turbulence-driven entrainment was investigated by \citet{lecoanet2018}, who found that the entrainment efficiency $e$ varies by less than $20\%$ between laminar (Re = 630) and turbulent (Re = 6300) thermals. This motivated us to utilize Turner's argument of buoyancy-controlled entrainment in addition to the thermal's vertical momentum equation to build a model for thermal dynamics which does not invoke turbulence or self-similarity. We derive simple expressions for the thermals' kinematic properties and their fractional entrainment rate $\epsilon$ and find close quantitative agreement with the values in direct numerical simulations. In particular, our expression for entrainment rate is consistent with the parameterization $\epsilon \sim B/w^2$, for Archimedean buoyancy $B$ and vertical velocity $w$. We also directly validate the role of buoyancy-driven entrainment by running simulations where gravity is turned off midway through a thermal's rise. The entrainment efficiency $e$ is observed to drop to less than 1/3 of its original value in both the laminar and turbulent cases when $g=0$, affirming the central role of buoyancy in entrainment in dry thermals.
    \end{abstract}
    \bigskip
    \bigskip
    \bigskip
    
    \section*{Funding Information}
    BM is supported by the NOAA Hollings Scholarship, DL is supported by a PCTS fellowship and a Lyman Spitzer Jr. fellowship, and NJ is supported by a Harry Hess fellowship from the Princeton Geoscience Department. Computations were conducted with support by the NASA High End Computing (HEC) program through the NASA Advanced Supercomputing (NAS) Division at Ames Research Center on Pleiades, as well as GFDL's computing cluster.
  \end{@twocolumnfalse}
]
\blfootnote{* Corresponding author address:
\href{mailto:brettmckim@gmail.com}{brettmckim@gmail.com}}
\blfootnote{$^\dagger$ Submitted to QJRMS}
\clearpage
\section{     Introduction} \label{sec: introduction}

Since the introduction of the entrainment hypothesis by \citet{morton1956}, the physical mechanism of entrainment has been thought to originate from turbulent eddies. The wide acceptance of this theory is due to its success in analyzing flows such as jets and plumes in a large variety of physical contexts \citep{turner1986}. The entrainment hypothesis states that the average inflow velocity across the edge of a turbulent flow is assumed to be proportional to a characteristic velocity. There are some flows however, where entrainment persists even in the laminar regime. For instance, ``thermals", or regions of isolated buoyant fluid thought to be the basic unit of convection \citep{romps2015,yano2014a}, were recently found to vary in entrainment by less than 20\% in laminar (Re = 630) and turbulent (Re = 6300) simulations in a dry atmosphere \citep{lecoanet2018}.

The observed expansion rates in thermals is much larger than that of plumes and is sensitive to the initial conditions \citep{turner1986,turner1973buoyancy}. According to \citet{turner1957}, these properties may be better understood if the thermal is modeled as a buoyant vortex ring. \citet{turner1957} shows why a buoyant ring must expand with time due to momentum conservation, and an illuminating, complementary perspective was provided by \citet{zhao2013} which shows that a ring's expansion is related to its baroclinicity. We refer the reader to \citet{zhao2013} for a more in-depth explanation.

Here, we combine the buoyant vortex ring argument of \citet{turner1957} with the thermal's vertical momentum equation by regarding the thermal as the region of fluid that moves coherently with the vortex ring (i.e. as the ellipsoidal region of fluid whose average velocity equals the gross translational velocity of the vortex ring itself). This aligns with the notion of a vortex ring ``bubble" \citep{Akhmetov2009,Shariff1992}, and we assume that the radius of the thermal $a$ is proportional to the radius of the vortex $R$, ie. $a = \xi R$. Utilizing this connection, we derive analytical expressions for the density, vertical velocity, height, and radius as functions of time. We then relate the radius to the height to get an expression for the fractional entrainment rate, $\epsilon \equiv \frac{d \log V}{dz} = e/a$, where $e$ is a proportionality constant called the ``entrainment efficiency" and $a$ is the radius of the thermal\footnote{This is equivalent to the mass flux $M$ formulation of entrainment $\epsilon \equiv \frac{d \log M}{dz}$ when the fluid is Boussinesq and detrainment is assumed negligible \citep{lecoanet2018}.}.  This is significant because our predictions require no tuning parameters to fit the data, unlike previous works, \citep{johari1992,turner1986,simpson1983b,escudier1973,simpson1969,turner1964b,turner1962,levine1959,morton1956}. We validate this model by comparing the predictions of the radius, height, density, vertical velocity, and fractional entrainment to direct numerical simulations of dry thermals.

Our model does not invoke turbulence, contrary to the substantial amount of literature which attributes entrainment in convection to turbulent eddies \citep{yano2014a,derooy2013,sherwood2013,derooy2010,Romps2010a,craig2008,ferrier1989,baker1984,escudier1973,turner1964b,morton1956}. In addition to validating our simple non-turbulent model against simulation data, we directly test the role of buoyancy by performing numerical simulations where gravity is removed midway through a thermal's rise. The thermals' observed entrainment rates decrease significantly. Our expressions and simulations seem to indicate that entrainment in dry thermals is predominantly a laminar process, consistent with \cite{sanchez1989} and contravening the original entrainment hypothesis. Furthermore, buoyancy-driven entrainment is distinct from the `dynamic' entrainment proposed in \citet{houghton1951} and used in \citet{morrison2017,derooy2010,ferrier1989,asai1967}, since dynamic entrainment is a consequence of \textit{positive} vertical acceleration, whereas our theory and simulations show \textit{negative} vertical acceleration. We conclude with a discussion of the implications of these findings on future studies of entrainment, given the importance of this process in understanding cumulus convection and climate sensitivity \citep{zhao2014,klocke2011}.

We begin by reviewing previous work on buoyant vortex rings in section \ref{sec: review} and discuss the relationship between a vortex ring and a thermal. In section \ref{sec: a new model} we introduce a new model which allows us to derive the characteristics of a thermal and its entrainment. In section \ref{sec: model verification} we verify the predictions made by the model and then affirm the role of buoyancy in entrainment. In section \ref{sec: discussion} we discuss second order contributions to entrainment and conclude by looking at future research directions.
\section{     The Physics of Buoyant Vortex Rings} \label{sec: review}
\subsection{Fundamental Fluid Mechanics} \label{sec: fundamental fluid mechanics}
% This section talks about the relationship between thermals and buoyant vortex rings, and then reviews previous work on buoyant vortex rings in subsections \ref{sec: preliminaries}, \ref{sec: conservation of momentum}, and \ref{sec: baroclinicity}. We will combine the insights in this review to create our new model of thermals in section \ref{sec: a new model}.

In our analysis of thermals we will use the Boussinesq approximation to the momentum equation, where density differences $\rho '$ in the flow are small compared to the constant background reference density $\Bar{\rho}$. We decompose the density as $\rho = \Bar{\rho} + \rho '$. The pressure can be decomposed similarly as $p = \Bar{p} + p'$ where $\Bar{p}$ is in hydrostatic balance with $\Bar{\rho}$, ie. $\nabla \Bar{p} = \Bar{\rho} \bm{g}$, where $\bm{g} = -g \bm{\hat{z}}$ is the gravitational acceleration. Formally, the Boussinesq approximation is valid when vertical length scales in the problem are smaller than the scale height. This approximation is helpful, as it eliminates the effect of adiabatic expansion of a fluid parcel as it rises. Therefore, the observed expansion of thermals will be purely due to the mechanical effect of entrainment. The integrated buoyancy force in the Boussinesq approximation is $F = - \int \rho' g \ dV$, where the integral is taking over the region of interest. The momentum equation is
$ \frac{D\bm{u}}{Dt} = \frac{1}{\Bar{\rho}} \big [ \rho '\bm{g} - \nabla{p'}\big ] + \nu \nabla ^2 \bm{u}$,
where $\bm{u}$ is the velocity field and $\nu$ is the kinematic viscosity.

\begin{figure}
  \includegraphics[width=\linewidth]{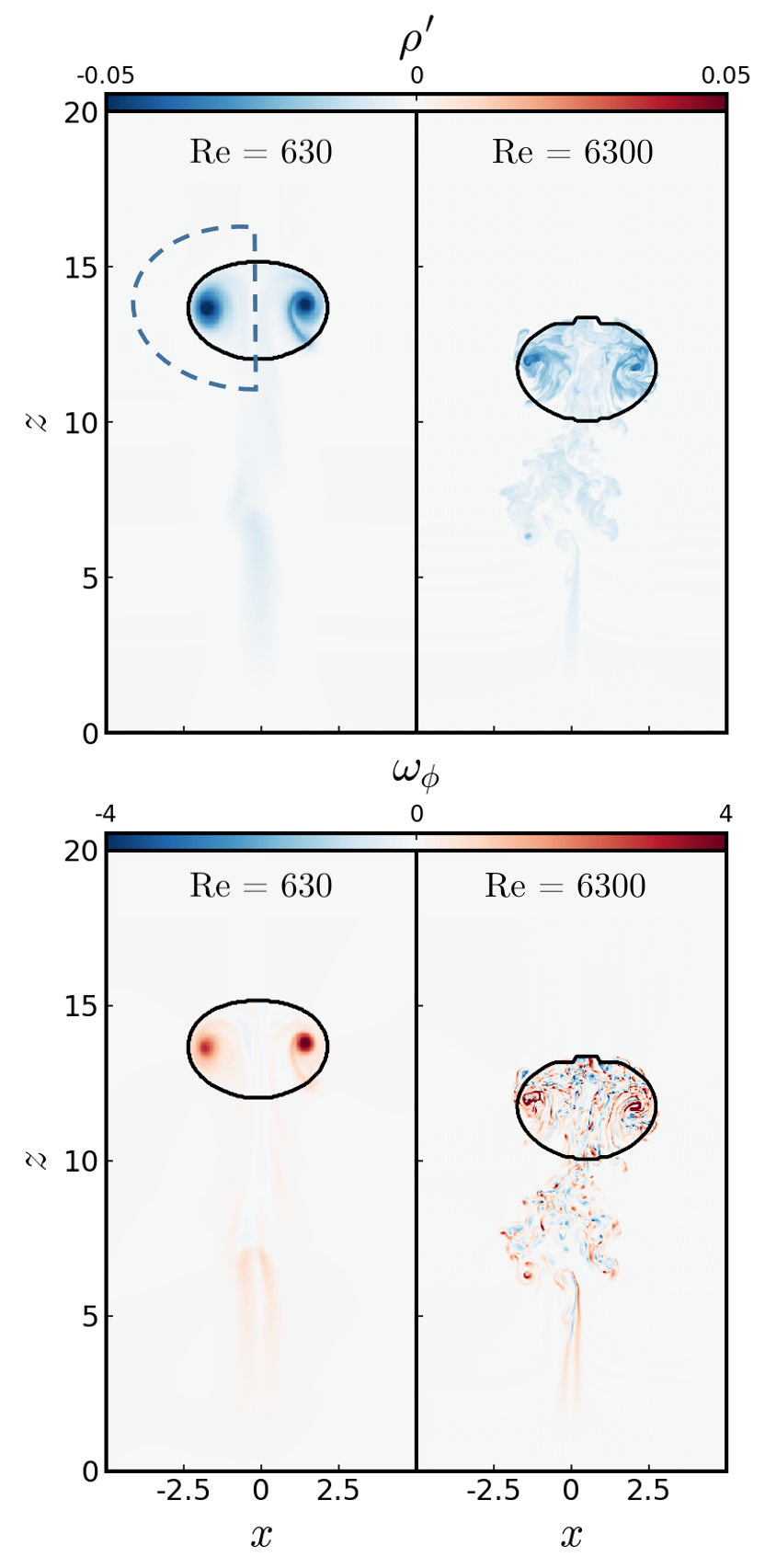}
  \caption{2D vertical slices at the thermal midpoint of the density perturbation (top) and vorticity (bottom) of two thermals with different Reynolds numbers at $\tau = 2.5$. The left (right) panel shows the Re=630 (Re=6300) simulations. The same color scale is used for both Re. The thermals originally start as a buoyant sphere (with noise added to break the symmetries of the problem), but then develop into a vortex ring that induces a flow structure that moves coherently with it. We identify this region as the thermal, and details of how it is computed is given \citet{lecoanet2018}, section 3. The contour of the thermal is plotted in black. Both the density perturbations and the vorticity become primarily concentrated within the core region of the vortex in the laminar thermal. This is also true in the turbulent thermal, but to a lesser extent due to vorticity fluctuations at many scales, characteristic of a turbulent fluid. The dashed blue line is an example of a circuit $\partial \mathcal{S}$ which passes through no buoyant fluid and can be used for Eqn. (\ref{Eq: Circulation}) and Eqn. (\ref{vortex momentum}).  }
  \label{fig:cross sections}
\end{figure}

\begin{figure}[h]
  \includegraphics[width=\linewidth]{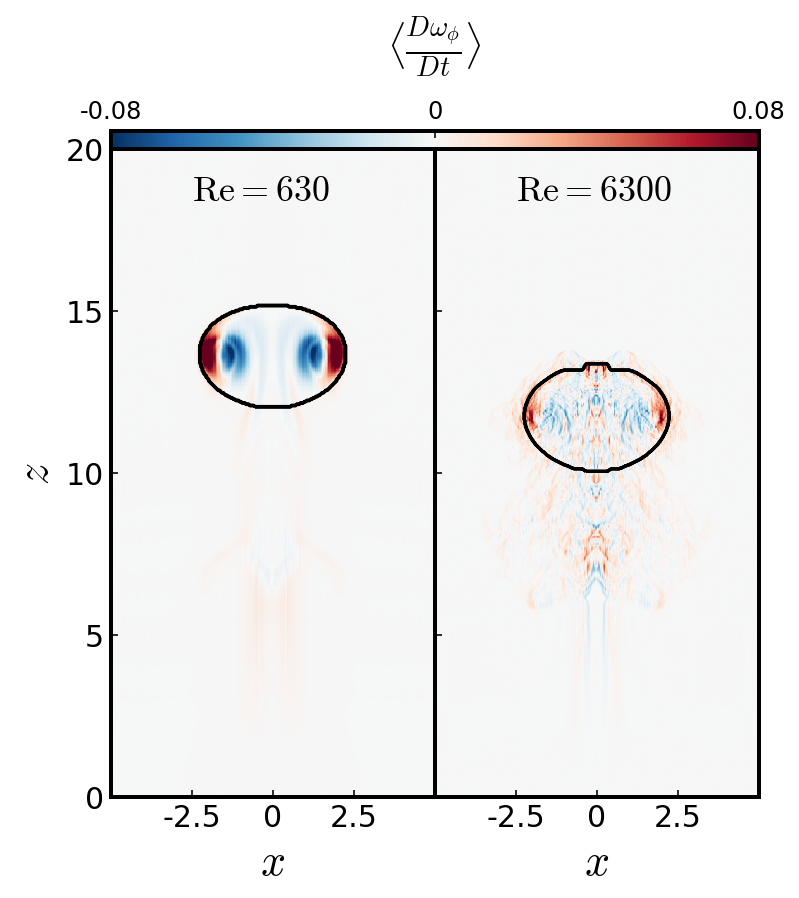}
  \caption{Azimuthally averaged vertical slices of the change in vorticity $\langle \frac{D\omega_{\phi}}{Dt} \rangle$ at $\tau = 2.5$. Angle brackets denote an azimuthal average has been taken, which is the reason for the symmetry of the figures across $x=0$. Radial gradients in the density field produce a torque (baroclinicity) that drives the continual destruction of vorticity in the center of the vortex and the continual creation vorticity in the outer region. This causes the center of rotation of the vortex to continually move outward \citep{zhao2013} and is the mechanical description of how dry thermals expand and therefore entrain as they rise.}
  \label{fig:zhao}
\end{figure}

When a spherical thermal is released from rest, it accrues vorticity from its baroclinicity and generates a vortex ring structure, where density perturbations $\rho '$ are largely confined to the vortex core that develops, Figure \ref{fig:cross sections}. Buoyancy is present only where there are density perturbations, therefore, the buoyancy resides mostly within the vortex core. In our model, we will assume that the buoyancy resides entirely within the vortex core. This assumption will be tested in Section 4. We introduce some basic aspects of vortex rings below.

\subsection{Vortex Ring Dynamics} \label{sec: preliminaries}
A vortex ring can be characterized by its kinematic properties such as circulation and impulse. The circulation of a vortex ring can be calculated along a circuit  $\partial \mathcal{S}$ passing through the center of the thermal's vortex ring and then returning through the ambient fluid, like in Figure \ref{fig:cross sections}. The integral can also be computed as an area integral of vorticity over the region bounded by the circuit, $\mathcal{S}$. For vortex rings, which are azimuthally symmetric, the vorticity is $\bm{\omega} = \omega_\phi \bm{\hat{\phi}}$ and this integral simplifies in cylindrical coordinates $(r,\phi,z)$, where the origin is located in the center of the vortex ring,

\begin{equation}
\Gamma \ \equiv \ \oint_{\partial \mathcal{S}} \bm{u} \cdot d \bm{l}\ = \ \int_\mathcal{S} \omega_\phi \ \ dr dz \ .
\label{Eq: Circulation}
\end{equation}
The impulse for incompressible fluids is defined as \citep{Shivamoggi2010,Akhmetov2009,batchelor2000,lamb1993hydrodynamics},

\begin{equation}
\bm{I} \ \equiv \ \frac{1}{2} \rho \int_\mathcal{V} \bm{x} \times \bm{\omega} \quad dV \ ,
\label{full vortex ring impulse}
\end{equation}
where $\bm{x}$ is the position vector and $\mathcal{V}$ is the entire domain. It can be interpreted as the time and volume integral of external forces that must be applied to a flow in order to generate the observed fluid motion from rest \citep{lamb1993hydrodynamics}. Therefore, internal forces such as pressure or viscosity do not come into play. Eqn. (\ref{full vortex ring impulse}) simplifies if we assume the flow is Boussinesq, that the radius of the core is much smaller than the radius of the ring $R$, and use the second part of Eqn. (\ref{Eq: Circulation}),

\begin{equation}
I_z \ = \ \pi \Bar{\rho} \ \Gamma R^{2} \ .
\label{vortex momentum}
\end{equation}
We only get an impulse in the z direction, which is a consequence of the azimuthal symmetry of the flow. We turn to the circulation equation to consider how circulation evolves in time, 

\begin{equation}
    \frac{d\Gamma}{dt} = \int_\mathcal{S} \bigg ( \frac{\nabla \rho ' \times \bm{g}}{\bar{\rho}} + \text{Re}^{-1} \nabla ^2 \bm{\omega} \bigg) \cdot d\bm{A}
    \label{eq: circ evolution}
\end{equation}

In our current analysis we will assume the Reynolds number is sufficiently high that the viscous effects can be neglected (though see section~\ref{sec: second order effects} for more discussion of viscous torques). If the circuit $\partial \mathcal{S}$ passes through no buoyant fluid, then the circulation is constant with time \citep{fohl1967}. This circuit will be possible once the vortex ring has formed, so we expect the circulation to initially increase with time as the thermal ``spins up" but then reach a constant value. We will verify this property of the circulation in section 4.

\subsection{Expansion by Conservation of Momentum} \label{sec: conservation of momentum}
\citet{turner1957}'s argument applies after the vortex ring reaches a constant circulation and can be summarized succinctly: Internal forces such as pressure or viscosity do not affect the impulse of the ring, so the change in impulse only depends on buoyancy.

\begin{equation}
\frac{dI_z}{dt} \ = \ \pi \Bar{\rho} \ \Gamma \frac{dR^{2}}{dt} \ = \ F \ > \ 0\ .
\label{buoyant vortex eq}
\end{equation}
$F$ and $\bar{\rho}$ are constants and therefore $R$ must increase with time, i.e. the thermal must entrain. For a generalization of this argument to the density-stratified case where $\bar{\rho}$ is not constant, see \cite{Anders2019}.

\subsection{Expansion by Baroclinicity} \label{sec: baroclinicity}
Although the argument in Eqn. \eqref{buoyant vortex eq} shows \textit{why} the radius of thermal increases with time, it is helpful to look at the vorticity equation to explain \textit{how}. We follow \citet{zhao2013} below. The vorticity equation in the Boussinesq approximation is,

\begin{equation}
\frac{D\bm{\omega}}{Dt} \ = \ 
 (\bm{\omega} \cdot \nabla)\bm{u} \ + \  \frac{\nabla \rho' \times \bm{g}}{\bar{\rho}} \ + \  \text{Re}^{-1} \nabla ^2 \bm{\omega} \ .
\label{vort eq}
\end{equation}
Once again ignoring viscous effects, in cylindrical coordinates we have

\begin{equation}
\frac{D\omega_\phi}{Dt} \ = \ u_r \frac{\omega_\phi}{r} \ + \   \frac{g}{\bar{\rho}}\frac{\partial \rho'}{\partial r}\ .
\label{cylindrical vort eq}
\end{equation}
The first term represents the intensification of vorticity due to stretching of vortex lines \citep{thorne2017modern}, but should disappear when integrated over the entire thermal because of the symmetry of the field, Figures \ref{fig:cross sections} and \ref{fig:zhao}. Equation (\ref{cylindrical vort eq}) then tells us that the vorticity evolution is dictated by the second term, baroclinicity, which depends on gradients in density perturbations. The gradients are always set up such that vorticity is constantly being created in the exterior of the ring and destroyed in the interior, as shown in Figure \ref{fig:zhao}, and thus the core region of the ring appears to continually move radially outward. As the ring expands, it entrains more fluid. This will be shown to be the dominant mechanism of entrainment in section \ref{sec: model verification}.

\section{     A New Model for Thermals} \label{sec: a new model}

\begin{figure}
  \includegraphics[width=\linewidth]{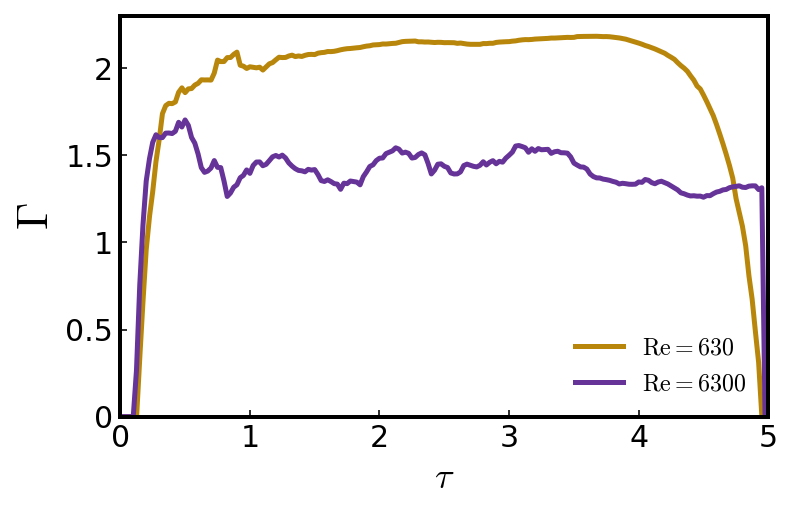}
  \caption{Verification of constant circulation $\Gamma$ for laminar (yellow) and turbulent (purple) simulations. For both cases, $\Gamma$ is roughly constant, which validates the approximation used to derive Eqn. (\ref{a eq}). The circulation also shows the thermals' spin up periods. We chose our timescale such that $\tau = 1$ is approximately the spin up time.}
  \label{fig:circulation}
\end{figure}

\begin{figure}
  \includegraphics[width=\linewidth]{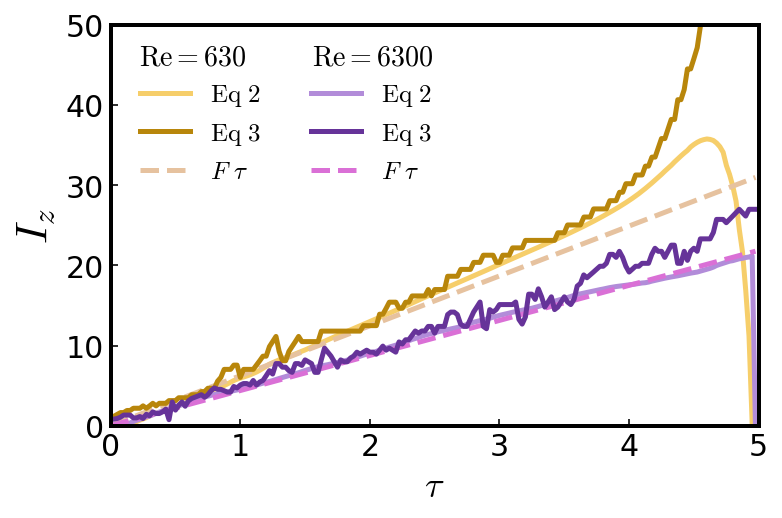}
  \caption{Verification of impulse expressions. The laminar thermal's impulse is shown in yellow and the turbulent's in purple. Equation (\ref{vortex momentum}) was derived assuming the radius of the core region of a vortex is small compared to its overall radius. We test the validity of this assumption by explicitly comparing it to the impulse computed by Eqn. (\ref{full vortex ring impulse}). We also assume integrated buoyancy is constant and therefore the impulse generated by it, $F \tau$, can be calculated too. We find that in both laminar and turbulent thermals, $F\tau$ tracks Eqn. (\ref{full vortex ring impulse}) closely.}
  \label{fig:momentum}
\end{figure}

Given the above picture for buoyancy-driven entrainment in thermals, we endeavour to build an analytical model for thermals which captures this picture, inspired by the formulation of \citet{escudier1973} who invoke the vertical momentum equation. In particular, we assume:

\begin{enumerate}
  \item{The flow is Boussinesq and begins as a buoyant sphere at $t=0$, but then fully spins up into a vortex ring at $t=t_o$ with radius $R_o$, height $z_o$, and vertical velocity $w_o$.}
  \item The impulse of a buoyant vortex ring is given by Eqn. (\ref{vortex momentum}), with circulation $\Gamma$ and background density $\bar{\rho}$ being constant.
  \item The only external force imparting an impulse is the integrated buoyancy $F$, which is constant because detrainment is assumed to be negligible \citep{lecoanet2018}. The Gross entrainment doesn't affect $F$ because the surrounding fluid has no mass anomaly. While $\langle \rho' \rangle$ will change and $V$ will change, their combination $\langle \rho' \rangle V$ will remain constant. For a thermal with volume $V=ma^3$,
  
  \begin{equation}
      F \ = \ m a^3 \big < \rho' \big > g \ = \  \text{constant} ,
    \label{buoyancy conservation}
  \end{equation}
  where $\big \langle \rho' \big \rangle$ is the average density anomaly of the thermal and $m$ is a constant for the volume since the thermal is no longer a sphere. $m$ is verified to be roughly constant in \citet{Anders2019}.
  \item The thermal is the region of fluid that moves coherently with the vortex ring and so the thermal's radius $a$ is proportional to the vortex radius $R$, ie. $a = \xi R$.
  \item The impulse of a thermal with volume $ma^3$ and vertical velocity $w$ can be written as \citep{Akhmetov2009},
  
  \begin{equation}
      I_z \ = \ ma^3\bar{\rho}(1+k)w \ ,
      \label{spherical thermal momentum eq }
  \end{equation}

where we parameterize virtual mass effects via a virtual mass coefficient $(1+k)$ \citep{batchelor2000}. For all ellipsoidal fluid geometries, the value of $k$ is the same as the solid body value of the same geometry, \citet{tarshish2018}.
\end{enumerate}
These assumptions will be discussed in more detail throughout the rest of the paper and will be tested along with the model predictions in section \ref{sec: model verification}.

We begin our analysis with the vortex momentum equation (\ref{vortex momentum}) which is equal to the impulse imparted by the integrated buoyancy, $Ft$. Assuming that $F$ is constant, we evaluate this equality at $t$ and $t_o$ and take their ratio to get the radius of the vortex ring as a function of time,

\begin{equation}
\begin{split}
R(t) \ &= \ R_{o} \sqrt{t/t_o} \ . \\
\end{split}
\label{r eq}
\end{equation} 
We utilize the proportionality of $a$ and $R$ and introduce a nondimensional time, $\tau = t/t_o$ to get the thermal's radius.

\begin{equation}
a(\tau) \ = \ a_{o}\sqrt{\tau} \ .
\label{a eq}
\end{equation}
Now consider the equation for integrated buoyancy conservation, (\ref{buoyancy conservation}). If we plug in Eqn. (\ref{a eq}) for $a$, we can solve for $\big <\rho' \big >$ explicitly as a function of time,

\begin{equation}
\big <\rho' \big >(\tau) \ = \ \frac{\big <\rho'_o \big >}{\tau^{3/2}} \ ,
\label{density anomaly eq}
\end{equation}
where $\langle \rho_o ' \rangle$ is the average density anomaly of the thermal at $t_o$. Now we turn to the Boussinesq momentum equation for a spherical thermal, (\ref{spherical thermal momentum eq }). In this formulation, a thermal is the region of fluid moving together with the vortex, so the impulse will be equal to the sum of the vortex momentum and the momentum created by virtual mass \citep{Akhmetov2009}. Evaluating Eqn.  (\ref{spherical thermal momentum eq }) at $t$ and $t_o$ and taking their ratio gives the velocity of the thermal as a function of time,

\begin{equation}
\begin{split}
w(\tau) \ &= \ w_o/\sqrt{\tau} \ .
\end{split}
\label{w eq}
\end{equation}
Now we integrate $w$ to get $z$.

\begin{equation}
\begin{split}
z(\tau) \ &= \ z_o \ + \ 2 w_o t_o \big (\sqrt{\tau} - 1 \big ) \ .
\end{split}
\label{z eq}
\end{equation}
Assuming that detrainment is negligible implies that integrated buoyancy is conserved. Therefore the fractional entrainment rate is also the fractional change in thermal volume with respect to height \citep{lecoanet2018}. Straight forward algebra shows $\epsilon = \frac{d\log V}{dz} = \frac{3}{a} \frac{da}{dz}$. Expressed as a function of the thermal's radius,

\begin{equation}
\begin{split}
\epsilon (a) \ &= \ \frac{3a_o}{2w_o t_o}\frac{1}{a} \ .
\end{split}
\label{eq: entrain_tau}
\end{equation}
This is one of the main results of this paper, along with having an analytical model for all the thermal's variables ($a, w, \langle \rho ' \rangle$) which does not invoke similarity or turbulence. Unlike \citet{escudier1973}, our approach does not require any additional empirical constants to determine the thermal's behavior. This is because \citet{escudier1973} use the turbulent entrainment assumption, $\frac{d(\rho a^3)}{dt} = 3\alpha \Bar{\rho} a^2 \lvert w \rvert$, ($\alpha$ is an unspecified constant tuned to match experiments), instead of Eqn. (\ref{vortex momentum}). They now have 4 unknowns ($a$,$w$,$\langle \rho' \rangle$,$\alpha$) and only 3 equations. In our case however, we employ (\ref{vortex momentum}) instead of an entrainment assumption to fully solve the system for $a$, $w$, and $\langle \rho ' \rangle$.

Our approach also gives a simple expression for the thermal's entrainment efficiency,

\begin{equation}
\begin{split}
e \ &= \ \frac{3a_o}{2w_o t_o} \ .
\end{split}
\label{eq: entrain efficiency}
\end{equation}
We can also use \eqref{vortex momentum}, \eqref{spherical thermal momentum eq }, and $I_z=Ft$ to  rewrite the efficiency in terms of the thermal's integrated buoyancy and circulation. This yields another key result, namely that $e$ is a constant dictated by the initial spin up of the thermal:

\begin{equation}
    e \ = \ \frac{3m(1+k)F\xi^4}{2\pi^2\bar{\rho}\Gamma^2} \ .
    \label{eq: efficiency rewritten}
\end{equation}
%Even in the laminar case, small differences in the initial density distribution will lead to different spin up conditions, affecting the vertical velocity and radius of the thermal. 
Thermals with different initial conditions will thus have different efficiencies, which perhaps explains the variance of efficiencies with respect to Re found in \citet{lecoanet2018}. Equation  (\ref{eq: efficiency rewritten}) suggests in particular that differences  in $e$ between the laminar and turbulent cases in that paper  may stem from differences in $\Gamma$ (Figure \ref{fig:circulation} below). Furthermore, variations in initial aspect ratio,  which  \citet{Lai2015} showed could explain the large range of thermal entrainment rates found in the literature, might also be understood via \eqref{eq: efficiency rewritten} in terms of their effect on $F$ and $\Gamma$.

\begin{figure}[h]
  \includegraphics[width=\linewidth]{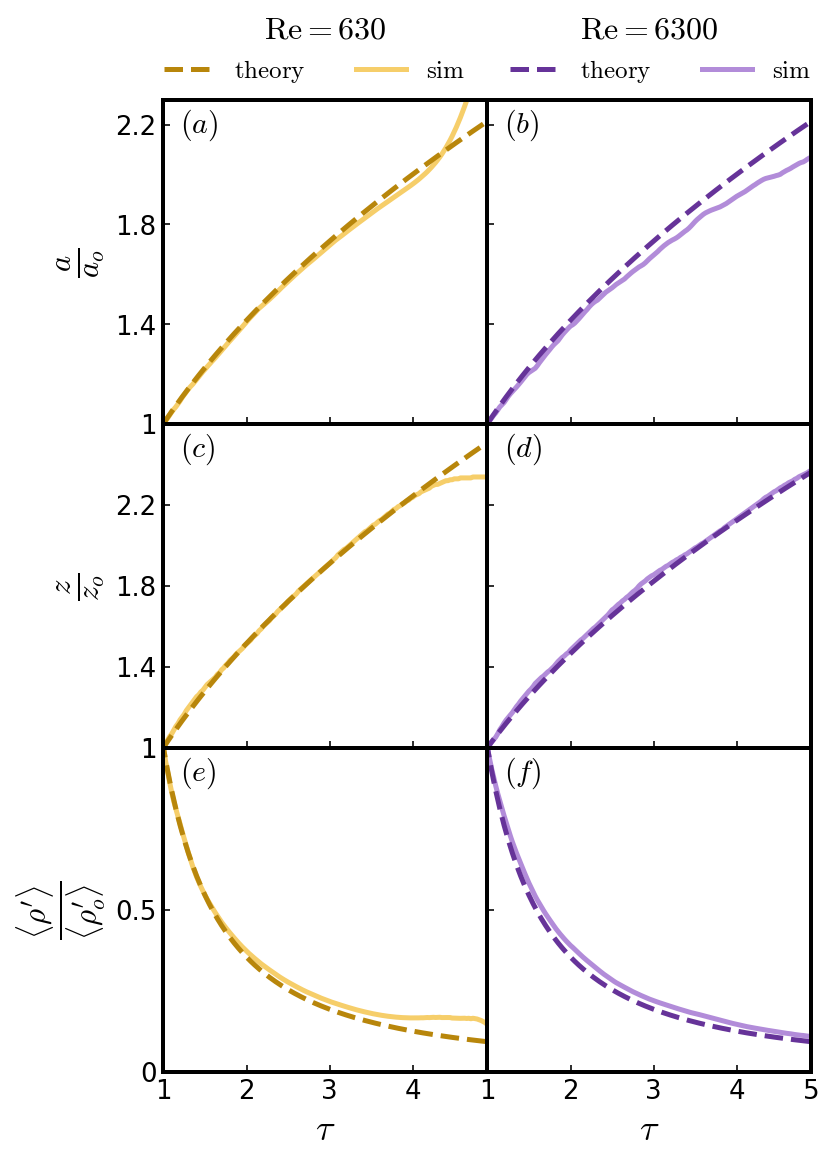}
  \caption{Verification of theory for thermal's (a)(b) radius $a$, (c)(d) cloud top height $z$, and (e)(f) average density perturbation $\langle \rho ' \rangle$. All values are normalized by their spin up value at $\tau = 1$. The predicted values from Eqns. (\ref{a eq}), (\ref{z eq}), and (\ref{density anomaly eq}) are the dashed lines and simulation values are the solid lines for the laminar (yellow) and turbulent (purple) cases. The deviations in the laminar data for $\tau >3.5$ is due to the thermal interacting with the top boundary. We find that both thermals exhibit power law behavior, consistent with \citet{lecoanet2018}.}
  \label{fig:scalings}
\end{figure}

\begin{figure}[h]
  \includegraphics[width=\linewidth]{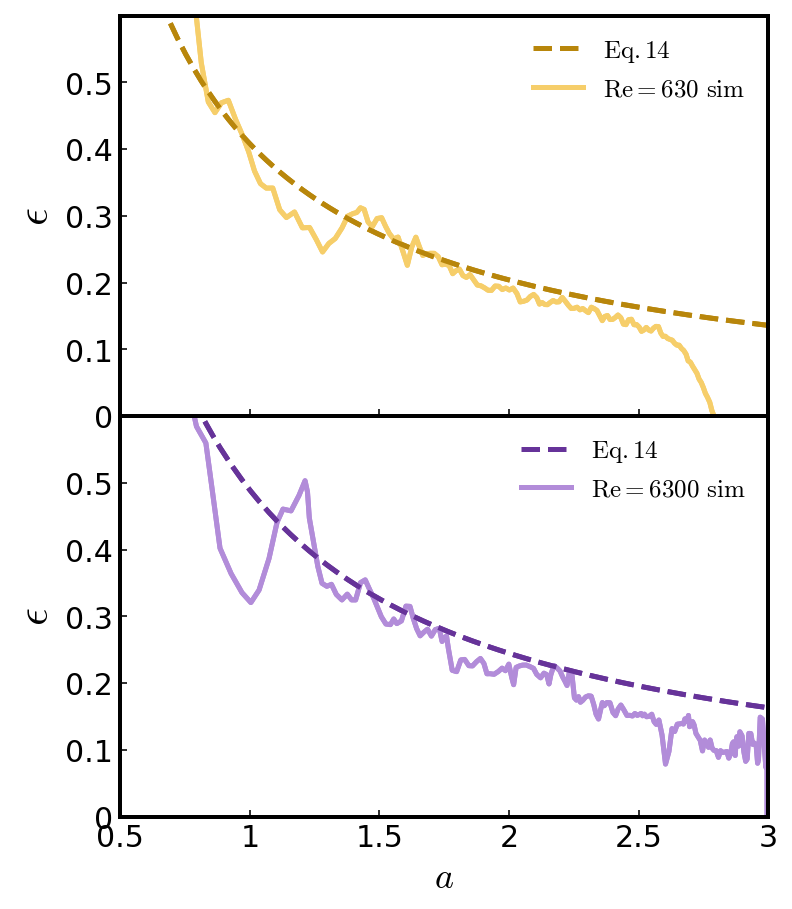}
  \caption{Verification of entrainment theory. Plotted is the fractional entrainment rate $\epsilon$ as a function of thermal radius $a$. Predicted values from Eqn. (\ref{eq: entrain_tau}) are dashed lines and simulation data are solid lines. Our model accurately predicts the entrainment rate in both laminar and turbulent simulations, although deviations are slightly larger in the turbulent case. Note the similarity in $\epsilon$ between laminar and turbulent simulations. Deviations occur for the laminar simulation at $a > 2.7$, because the thermal starts to interact with the top boundary.}
  \label{fig:entrainment} 
\end{figure}

Another important aspect of \eqref{eq: efficiency rewritten} is that it embodies the well-known scaling  $e\sim \frac{F}{\bar{\rho}\Gamma^2}$ \citep[e.g.][]{Lai2015,Bond2010,fohl1967,turner1957}. If we apply this scaling to the entrainment rate $\epsilon$ and write it using $\Gamma\sim wa$ and $F\sim \bar{\rho}B a^3$ (where $B=g\langle\rho'\rangle/\bar{\rho}$ is the average Archimedean buoyancy), we find
\begin{equation}
    \epsilon \sim \frac{B}{w^2} \ .
    \label{eq_eps_scaling}
\end{equation}
Though this scaling is not often applied to atmospheric convection, it was recently employed (on largely dimensional grounds) in the parameterization of \cite{tan2018}, and was also used in \cite{Gregory2001}. Equation \eqref{eq: efficiency rewritten} provides a precise foundation for \eqref{eq_eps_scaling}, at least in the idealized dry case. 

%====================%
% Model Verification %
%====================%
\section{     Verification of the Model} \label{sec: model verification}
\subsection{Simulation Setup} \label{sec: sim setup}

We analyze the direct numerical simulations of thermals found in \citet{lecoanet2018}\footnote{Although they ran an ensemble of simulations, we restrict our analysis to the simulations with the fifth initial condition.}. We outline the simulation setup below, but for more details (including the scalings used for nondimensionalization), see section 2 of \citet{lecoanet2018}. Simulations are run with Dedalus, an open source pseudo-spectral framework \citep{2019arXiv190510388B}. We solve the non-dimensionalized Boussinesq equations:

\begin{subequations}
    \begin{align}
        \label{eq: sim equationsa}
        \partial_t \bm{u}  +   \nabla p  -   \mathrm{Re}^{-1} \nabla ^2 \bm{u}  +   \rho ' \bm{e_z}  \ &= \  - \bm{u} \cdot \nabla \bm{u} \\
        \label{eq: sim equationsb}
        \partial_t \rho ' \ - \ \mathrm{Pr}^{-1} \mathrm{Re}^{-1} \nabla ^2 \rho' \ &= \ - \bm{u} \cdot \nabla \rho ' \\
        \nabla \cdot \bm{u} \ &= \  0
        \label{eq: sim equationsc}
    \end{align}
\end{subequations}
After non-dimensionalization, the simulations are entirely described by the initial condition, the Reynolds number and the Prandtl number. We take $ \text{Pr} = 1$ for all simulations, $\text{Re} = 630$ for laminar runs, and $\text{Re} = 6300$ for turbulent runs. 

We initialize the thermal with a spherical density anomaly $\rho'$ of magnitude negative $1$ and diameter $L_{th}$, plus a noise field to break symmetries in the problem. The spherical anomaly is placed near the bottom of a 3D domain with height $20 L_{th}$ and horizontal lengths $10 L_{th}$. The simulation time ranges from $t \in [0,63.2]$, long enough for the thermal to approach or hit the top boundary. We utilize the thermal tracking algorithm described in section 3 of \citet{lecoanet2018}. This defines the thermal to be the axisymmetric volume whose averaged vertical velocity matches the velocity of the thermal's top, where the thermal top is defined by a $\rho'$ threshold. Utilizing this tracking method, we can calculate the height, radius, velocity, and other dynamic quantities of the thermal. These measurements will serve as the test for the equations developed in section \ref{sec: a new model}. All comparisons use a spin up time of $t_o = 4\sqrt{10} \approx 12.6$, which is determined by looking at when the thermal's circulation reaches a constant value, Figure \ref{fig:circulation}. All of the code used to analyze the simulations in this work can be found online in a Github repository (github.com/mckimb/buoyant\_entrainment).

\subsection{Comparison to Simulations} \label{sec: sim comparison}

The equation for the radius of the vortex ring (\ref{r eq}) depended on an assumption of constant circulation $\Gamma$ within the thermal. While the constancy of circulation in Boussinesq thermals was confirmed in tank experiments in \citet{zhao2013}, we thought it worthwhile to also confirm this property in our simulations. We obtain the circulation by first calculating the azimuthal vorticity field, selectively choosing only the data points which lie within the thermal boundary, and then azimuthally averaging the data. We then take an area integral over the thermal's domain which is equivalent to the integral in the second part of Eqn. (\ref{Eq: Circulation}). The resulting $\Gamma$ is fairly constant in the laminar and turbulent simulations  (Figure \ref{fig:circulation}). The laminar thermal's circulation does increase slightly over time, which may be due to small baroclinic contributions in Eqn. (\ref{eq: circ evolution}). Variations in the turbulent case are larger, which can be attributed to buoyant fluid not contained entirely within the core of the vortex ring and small detrainment events which occur throughout the thermal's rise.

Simplifying Eqn. (\ref{full vortex ring impulse}) to Eqn. (\ref{vortex momentum}) made relating the impulse to the thermal's dynamic variables analytically tractable. In doing so, we assume the integrated buoyancy is constant and imparts an impulse $F \tau$, and that the radius of the core region of the vortex is much smaller than the radius of the entire vortex. Visually inspecting Figure \ref{fig:cross sections} casts doubt on the validity of this assumption as the core region is clearly finite. Despite this, we find that the approximations closely follow the actual impulse for both laminar and turbulent simulations (Figure \ref{fig:momentum}), verifying their validity. Deviations in the laminar simulation grow for $\tau > 3.5$ because the thermal begins to interact with the top boundary.

We now test the predictions for the thermals' characteristics, $a(\tau)$, $\langle \rho' \rangle (\tau)$, and $z(\tau)$ made in section \ref{sec: a new model} and plotted in Figure \ref{fig:scalings}. All plots are normalized by their spin up value at $\tau = 1$. We find that Eqns. (\ref{a eq}), (\ref{density anomaly eq}), and (\ref{z eq}) agree quite closely with the simulations. Deviations occur for the laminar simulation at $\tau > 3.5$, because the thermal starts to interact with the top boundary.

We verify the prediction Eqn. (\ref{eq: entrain_tau}) of the fractional entrainment rate, $\epsilon(a)$ in Figure \ref{fig:entrainment}. The predictions match the directly measured entrainment closely. Deviations in the laminar case occur for $\tau > 3.5$ because the thermal begins to interact with the top boundary of the simulation. We find that the turbulent prediction overestimates the actual entrainment, but the reasons for this are unclear, given how well the approximations in the theory hold up.

\subsection{Mechanism Denial of Buoyancy-Driven Entrainment} \label{label: mechanism affirmation}

\begin{figure}
  \includegraphics[width=\linewidth]{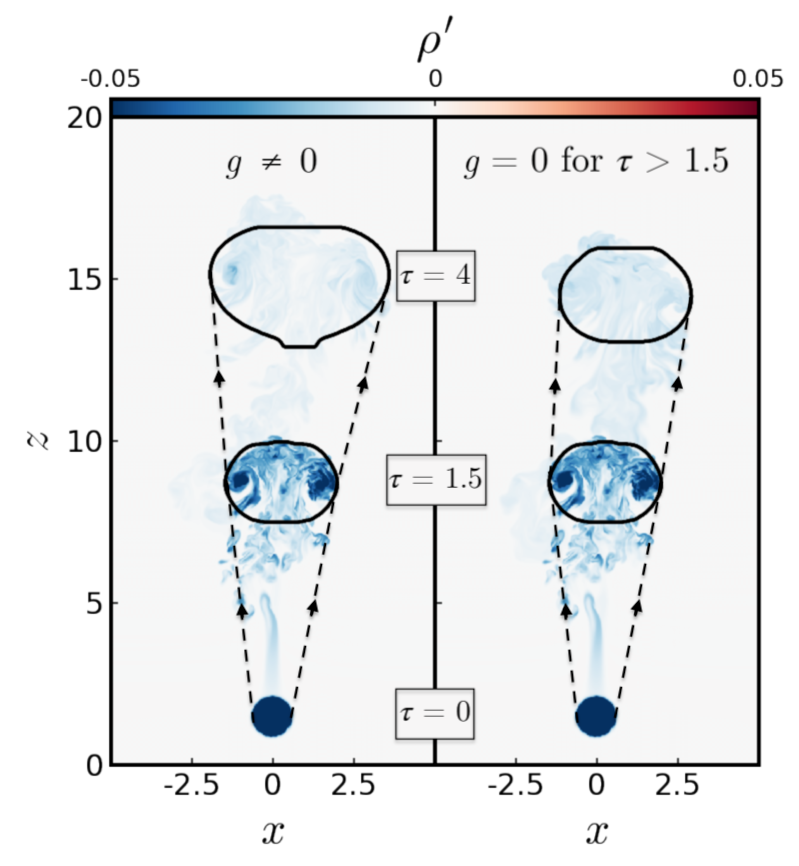}
  \caption{2D vertical slices at the thermal midpoint of the density perturbation at $\tau = 0$, $\tau = 1.5$, and $\tau = 4$. The left panel shows the unaltered Re=6300 simulation, where the thermal's evolution is self similar. The right panel shows the mechanism denial simulation which starts with the same initial conditions, but then is altered by removing gravity when $\tau > 1.5$. The thermal continues to rise because of its impulse, but it no longer expands appreciably because the buoyancy-driven mechanism of entrainment has been removed. Note that the density perturbation field reduces to a passive scalar when $g=0$ because it is no longer present in the governing equations.}
  \label{fig: g_nog}
\end{figure}

To verify that buoyancy (not turbulence) is the dominant source of entrainment, we ran both laminar and turbulent simulations in which the gravitational force is set to zero partway through the simulation. We turn off gravity at $\tau=1.5$, after the thermal has already spun-up into a vortex ring. The sudden removal of gravity in the simulation skews the balance of forces in the fluid, causing the thermal to assume a new trajectory which is no longer self-similar (Figure \ref{fig: g_nog}). Absent entrainment, the vortex ring should rise with constant velocity. In the laminar case, the height indeed increases linearly with time. However, in the turbulent case, we find the height still increases as the square-root of time, similar to buoyant thermals. To calculate the thermal volume, we respectively use linear and square-root time dependence to approximate the height (see \citet{lecoanet2018} for more details about the thermal tracker). In both cases, the thermal no longer grows in size appreciably compared to the unmodified simulations, as shown in Figure \ref{fig: g_nog} below.

Calculating entrainment accordingly, we see that in both the laminar and turbulent cases, $e$ drops significantly, Figure \ref{fig: g=0}. Without buoyancy, the observed entrainment efficiency drops dramatically, giving the most direct evidence of the central role of buoyancy in entrainment. This nicely complements the \textit{mechanism affirmation} experiments of \citet{lecoanet2018} who kept buoyancy and instead removed turbulence to show that the Reynolds number had little effect on the measured entrainment rate.

\section{     Discussion} \label{sec: discussion}
\subsection{Second Order Effects} \label{sec: second order effects}

\begin{figure}
  \includegraphics[width=\linewidth]{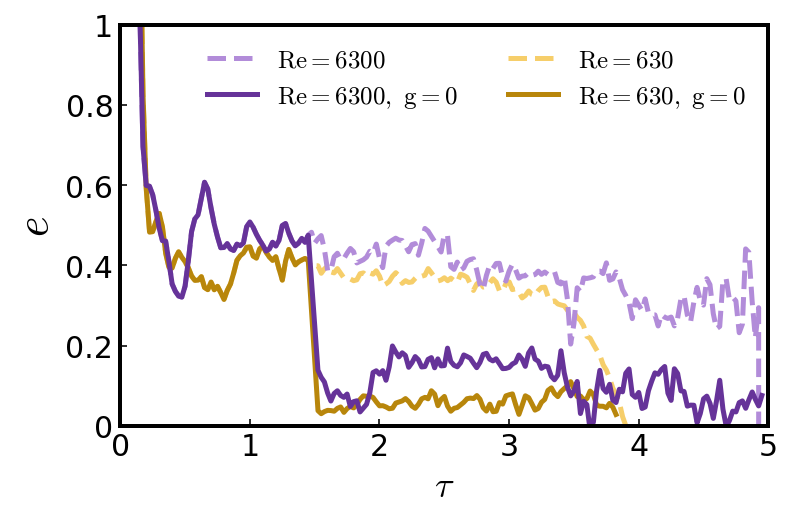}
  \caption{Confirmation of buoyancy-driven entrainment. Plotted is the entrainment efficiency $e$ as a function of time. We directly test the idea of buoyancy-driven entrainment by setting $g=0$ at $\tau = 1.5$, sufficiently after the thermals have spun up. The original simulations are dashed lines, and the simulations with where gravity is removed midway through are solid lines. We find in both laminar (yellow) and turbulent (purple) thermals the entrainment efficiency sharply drops off to less than 1/3 of their original values, giving the most direct evidence of the central role of buoyancy in entrainment. The residual entrainment that remains can be attributed to viscous effects, see section \ref{sec: second order effects}. Deviations occur for the laminar simulations at $\tau > 3.5$, because the thermal starts to interact with the top boundary.}
  \label{fig: g=0}
\end{figure}

Buoyancy is essential and is the leading order effect in entrainment in dry thermals. However, even after removing gravity from the simulations, both the laminar and turbulent thermals continue to entrain a small amount (Figure \ref{fig: g=0}). Note that there are no baroclinic torques without gravity.
Equation (\ref{vortex momentum}) hence implies $R \sim 1/\sqrt{\Gamma}$ so an increase in $R$ suggests the circulation is decreasing with time. 
The residual entrainment appears to depend on the level of turbulence of the thermal, as the entrainment in the $\mathrm{Re}=6300$ simulation is roughly double the entrainment in the $\mathrm{Re}=630$ simulation.
Furthermore, we find the laminar thermal's height increases roughly linearly with time, whereas the turbulent thermal's height follows a square root time dependence.

In the absence of gravity, the circulation evolves according to

\begin{equation}
    \frac{d\Gamma}{dt} \ = \ \int_\mathcal{S}  \text{Re}^{-1} \nabla ^2 \bm{\omega} \cdot d\bm{A} \ \lesssim 0.
    \label{eq: Circulation change}
\end{equation}
We hypothesize the entrainment seen in Figure~\ref{fig: g=0} is \textit{viscous entrainment}.
It may then be puzzling that the turbulent thermal (with lower viscosity) entrains more than the laminar thermal (with higher viscosity).
To understand this, we will now estimate the size of the integrand in Eqn. (\ref{eq: Circulation change}) in both laminar and turbulent cases.

In the laminar thermal, the vorticity is largest on the lengthscale of the vortex radius $R$, so we can estimate $\omega_\phi \sim w/R$ and $\nabla\sim R^{-1}$. Then the integrand scales like

\begin{equation}
    \quad\quad
    \text{Re}^{-1} \frac{w}{R^3} \quad\quad \text{(laminar)}.
\end{equation}
However, in the turbulent case, we see the vorticity is largest on small scales (Figure~\ref{fig:cross sections}).
For turbulent flows, the enstrophy is typically maximum near the viscous scale, $L_v\sim R \, \mathrm{Re}^{-3/4}$ \citep{thorne2017modern}.
We then estimate $\omega_\phi \sim w/L_v$ and $\nabla\sim L_v^{-1}$, so the integrand scales like

\begin{equation}
    \quad\quad
    \mathrm{Re}^{5/4} \frac{w}{R^3} \quad\quad \text{(turbulent)}.
\end{equation}
This illustrates that the viscous torques may change the circulation more in turbulent thermals than laminar thermals! This is consistent with the substantially larger entrainment in turbulent thermals than laminar thermals when there is \textit{no} gravity (Figure \ref{fig: g=0}). Hence, absent gravity, we find there is a small amount of residual entrainment, which is turbulent in origin. These effects are small for buoyant thermals in the presence of gravity.

One subtlety in the argument above is that the integrand in the laminar case is everywhere negative, whereas in the turbulent case, it has both negative and positive contributions, leading to substantial cancellations. We expect that cancellations in the turbulent case will make the net change in circulation due to viscous torques independent of Reynolds number, rather than increasing with Reynolds number.
We also note that the slight \textit{increase} in circulation for the laminar simulation including gravity (Figure~\ref{fig:circulation}) is likely due to baroclinic torques, given that the viscous torques should decrease the circulation.

\subsection{Conclusion}
In this paper, we developed a simple theory for thermals based on the vertical momentum equation (\ref{spherical thermal momentum eq }) and the kinematic constraint (\ref{vortex momentum}). This allowed us to derive a complete theory for dry thermals without any additional, empirical constants. We also demonstrated that dry thermals entrain primarily through a buoyancy-driven processes by setting $g=0$ midway through the simulation runs. The entrainment is seen to drop off sharply and the small, residual entrainment that remains can be attributed to viscous dissipation that depends on the Reynolds number of the thermal. This is in contrast to many other explanations of entrainment in convection which rely on a primarily turbulent or stochastic entrainment hypothesis \citep{romps2010b,escudier1973,morton1956}. 

Our theory also resulted in the explicit expression \eqref{eq: efficiency rewritten} for the entrainment efficiency, which captures the well-known $F/\Gamma^2$ scaling for $e$ and is consistent with parameterizations of gross entrainment as proportional to $B/w^2$, as in \cite{tan2018} and \cite{Gregory2001}.
The verification of the $B/w^2$ scaling here and the $1/a$ scaling in \cite{lecoanet2018} might also explain why these scalings outperformed others in the moist convection study of \cite{Hernandez2018}.

To truly bridge the gap, however, between our idealized dry thermals and real world cumulus convection, future work should  develop a new suite of simulations for moist thermals, perhaps with a simplified model of condensation as in \citet{Vallis2018}. The current theory will likely need modification because the assumption of constant integrated buoyancy breaks down in moist thermals, which can generate buoyancy in their center due to latent heat release \citep{morrison2018,romps2015}. As for our Boussinesq approximation, the framework presented in this paper has already been generalized to anelastic and compressible thermals in \citet{Anders2019}. They study thermals in the context of solar convection, and the excellent agreement between theory and simulation in their more general paradigm suggests that our framework might also be amenable to moist thermals. A simple model for the moist case would provide a foundation on which a thorough understanding of cumulus convection may be built, bridging the gap between simulation and theory in climate and atmospheric modeling \citep{jeevanjee2017a,Held2005}.

\section*{Acknowledgements}
BM is supported by the NOAA Hollings Scholarship, DL is supported by a PCTS fellowship and a Lyman Spitzer Jr. fellowship, and NJ is supported by a Harry Hess fellowship from the Princeton Geoscience Department. Computations were conducted with support by the NASA High End Computing (HEC) program through the NASA Advanced Supercomputing (NAS) Division at Ames Research Center on Pleiades, as well as GFDL's computing cluster. We would like to thank Leo Donner and Nathaniel Tarshish for helpful discussions at many points throughout the research process and for Spencer Clark's computer support.

\bibliographystyle{apalike}
\bibliography{references}

\end{document}